\let\mymarginpar\marginpar

\documentclass[a4paper,envcountsame]{llncs}

\usepackage{times}

\usepackage{amsfonts}
\usepackage{amsmath}
\usepackage{amssymb}
\usepackage{ntheorem}
\theorembodyfont{\upshape}
\newtheorem{Definition}{Definition}

\usepackage{graphicx}
\usepackage{wrapfig}
\usepackage{hyperref}

\usepackage{chngpage}

\usepackage[T1]{fontenc}
\usepackage[scaled=0.81]{beramono} 
\usepackage{eiffel}
\newcommand{\e}[1]{\mbox{\lstinline|#1|}}

\newcommand{\vb}[1]{\mbox{\lstinline[language=Java]|#1|}}

\usepackage{paralist}

\renewcommand{\And}{\wedge}
\newcommand{\Or}{\vee}
\newcommand{\Implies}{\Rightarrow}

\newcommand{\reads}[1]{\mathsf{reads}(#1)}
\newcommand{\writes}[1]{\mathsf{writes}(#1)}
\newcommand{\guard}[1]{\mathsf{guard}(#1)}
\newcommand{\Heap}{\mathsf{H}}

\newcommand{\byes}{$\oplus$}
\newcommand{\yes}{$+$}
\newcommand{\no}{$-$}

\usepackage{subcaption}
\usepackage[table]{xcolor}
\usepackage{threeparttable}

\usepackage{tikz}
  \usetikzlibrary{shapes,arrows,positioning,calc}

\tikzset{
  object/.style={circle, minimum size=3mm,thick,draw,font=\scriptsize},
  node distance=1.5mm and 7mm,
  every edge/.style={-latex,thick,draw},
  every label/.style={font=\scriptsize},
  level distance=5mm, level/.style={sibling distance=12mm}
}

\usepackage{xspace}

\definecolor{shadingcolor}{rgb}{.85,.85,.85}

\makeatletter
\renewcommand*{\thetable}{\arabic{table}}
\renewcommand*{\thefigure}{\arabic{figure}}
\let\c@table\c@figure
\makeatother

\newcommand{\fakepar}[1]{\vskip 1pt\textbf{#1.}$\ $}

\newif\ifdraft\draftfalse

\ifdraft
  \newcommand\rmark[1]{\mymarginpar{\raggedright\hbadness=10000\tiny\it #1\par}}
\else
  \newcommand\rmark[1] {}
\fi

\ifdraft\overfullrule3pt\fi

\newif\ifextended\extendedtrue

\newcommand{\comment}[2]{\rmark{\textbf{{#1}}: {#2}}}
\newcommand{\caf}[1]{\comment{\textup{caf}}{#1}}

\usepackage[capitalize]{cleveref}
  \crefname{section}{Sect.}{Sect.}
  \Crefname{section}{Section}{Sections}
  \crefname{figure}{Fig.}{Fig.}
  \Crefname{figure}{Figure}{Figures}
  \crefname{listing}{Listing}{Listings}
  \Crefname{listing}{Listing}{Listings}
  \crefname{table}{Tab.}{Tab.}
  \Crefname{table}{Table}{Table}
  \crefname{equation}{}{}
  \crefname{Definition}{Def.}{Def.}
  \Crefname{Definition}{Definition}{Definitions}
  \crefname{lemma}{Lem.}{Lem.}
\usepackage{threeparttable}

\begin{document}

\tolerance=5000

\setlength{\intextsep}{7pt}
\setlength{\textfloatsep}{5pt}

\title{Flexible Invariants Through Semantic Collaboration%
\thanks{Work partially supported by SNF grants LSAT/200020-134974, ASII/200021-134976, and FullContracts/200021-137931; and by Hasler-Stiftung grant \#2327.}%
}

\author{Nadia Polikarpova \and Julian Tschannen \and Carlo A.\ Furia \and Bertrand Meyer}
\institute{%
Department of Computer Science, ETH Zurich, Switzerland\\
\email{firstname.lastname@inf.ethz.ch}
}

\maketitle

\begin{abstract}
Modular reasoning about class invariants is challenging in the presence of collaborating objects that need to maintain global consistency.
This paper presents \emph{semantic collaboration}: a novel methodology to specify and reason about class invariants of sequential object-oriented programs, which 
models dependencies between collaborating objects by semantic means. 
Combined with a simple ownership mechanism and useful default schemes, 
semantic collaboration achieves the flexibility necessary to reason about complicated inter-object dependencies
but requires limited annotation burden when applied to standard specification patterns.
The methodology is implemented in AutoProof, our program verifier for the Eiffel programming language (but it is applicable to any language supporting some form of representation invariants).
An evaluation on several challenge problems proposed in the literature
demonstrates that it can handle a variety of idiomatic collaboration patterns,
and is more widely applicable than the existing invariant methodologies.
\end{abstract}

\section{The Perks and Pitfalls of Invariants}
\label{sec:introduction}

Class invariants\footnote{Also known under the names ``object invariants'' or ``representation invariants''.} are here to stay~\cite{Need}---%
even with their tricky semantics in the presence of callbacks and inter-object dependencies, which make reasoning so challenging~\cite{Parkinson07}.
The main reason behind their widespread adoption is that
they formalize the notion of \emph{consistent} class instance,
which is inherent in object-orientated programming,
and thus naturally present when reasoning, even informally, about program behavior.

The distinguishing characteristic of invariant-based reasoning is \emph{stability}:
it should be impossible for an operation $m$ to violate the invariant of an object $o$ without modifying $o$ itself.
Stability promotes information hiding and simplifies client reasoning about preservation of consistency:
without invariants a client would need to know which other objects $o$'s consistency depends on,
while with invariants it is sufficient that it checks whether $m$ modifies $o$---a piece of information normally available as part of $m$'s specification. 
The goal of an \emph{invariant methodology} (also called \emph{protocol}) is thus to achieve stability even in the presence of inter-object dependencies---%
where the consistency of $o$ depends on the state of other objects, possibly recursively or in a circular fashion (see \cref{sec:examples} for concrete examples).

The numerous methodologies introduced over the last decade, which we review in \cref{sec:existing}, successfully relieve several difficulties involved in reasoning with invariants; but there is still room for improvement in terms of flexibility, usability, and automated tool support.
In this paper, we present \emph{semantic collaboration} (SC): a novel methodology for specifying and reasoning about invariants in the presence of inter-object dependencies that combines flexibility and usability and is implemented in a program verifier.

A standard approach to inter-object invariants is based on the notion of \emph{ownership},
which has been deployed successfully in several invariant methodologies~\cite{Barnett04,Dynamic,Muller06} and is available in tools such as Spec\#~\cite{SpecSharp} and VCC~\cite{VCC}.
Under this model, an invariant of an object $o$ only depends on the state of the objects explicitly owned by $o$.
Ownership is congenial to object-orientation because it supports a strong notion of encapsulation; however, not all inter-object relationships are hierarchical and hence reducible to ownership.
Multiple objects may also \emph{collaborate} as equals, mindful of each other's consistency;
a prototypical example is the Observer pattern~\cite{GOF-book} (see \cref{sec:examples}).

Semantic collaboration 
naturally complements ownership to accommodate invariant patterns involving collaborating objects.
Most existing methodologies support collaboration through dedicated specification constructs and syntactic restrictions on invariants~\cite{Dynamic,Friends,Middelkoop08,Considerate};
such disciplines tend to work only for certain classes of problems.
In contrast, SC relies on standard specification constructs---ghost state and invariants---%
to keep track of inter-object dependencies,
and imposes \emph{semantic} conditions on class invariant representations.
Its approach builds upon the philosophy of \emph{locally-checked invariants} (LCI)~\cite{Cohen10}: a low-level verification method based on two-state invariants.
LCI has served as a basis for other specialized, user- and automation-friendly methodologies
for ownership and shared-memory concurrency.
SC can be viewed as an improved specialization of LCI for object collaboration. 
To further improve usability, SC comprises useful ``defaults'', which characterize typical specification patterns.
\ifextended
As we argue in \cref{sec:evaluation} based on several challenge problems, the defaults significantly reduce the annotation burden without sacrificing flexibility in the general case.
\fi

We implemented SC as part of AutoProof, our automated verifier for the Eiffel object-oriented programming language.
The implementation provides more concrete evidence of the advantages of SC compared to other methodologies to specify collaborating objects (e.g., \cite{Friends,History,Considerate,Middelkoop08} all of which currently lack tool support).
\ifextended 

\fakepar{Outline and contributions}
The presentation is based on examples of non-hie\-rar\-chi\-cal object structures, customarily used in the literature.
\cref{sec:examples} presents the examples and the challenges they embody; and \cref{sec:existing} discusses the approaches taken by main existing invariant methodologies.
\cref{sec:methodology} introduces SC, demonstrates its application to the running examples, and outlines a soundness proof.
\cref{sec:evaluation} evaluates both SC and existing protocols on an extended set of examples,
including challenge problems from the SAVCBS workshop series~\cite{SAVCBS}.
\fi
The evaluation demonstrates that SC is the only methodology that supports
\begin{inparaenum}[(\itshape a)]
\item \label{contrib1} collaboration with unknown classes, while preserving stability, and
\item \label{contrib2} invariants depending on unbounded sets of objects, possibly unreachable in the heap.
\end{inparaenum}
The collection of problems of \cref{sec:evaluation}---%
available at~\cite{SC} together with our solutions---%
could serve as a benchmark to evaluate invariant methodologies for non-hierarchical object structures.
\ifextended
The website~\cite{SC} also gives access to AutoProof through a web interface.
\fi

\section{Motivating Examples: Observers and Iterators}
\label{sec:examples}

The \emph{Observer} and \emph{Iterator} design patterns are widely used programming idioms~\cite{GOF-book},
where multiple objects depend on one another and need to maintain a global invariant.
Their interaction schemes epitomize cases of inter-object dependencies that ownership cannot easily describe; 
therefore, we use them as illustrative examples throughout the paper, following in the footsteps of much related work~\cite{History,Parkinson07,Middelkoop08}.

\begin{figure}[!tb]
\lstset{basicstyle=\scriptsize,xleftmargin=8mm}
\centering
\ifextended 
\begin{tabular}{p{.49\textwidth} p{.49\textwidth}}
\begin{lstlisting}
class SUBJECT

  value: INTEGER
  subscribers: LIST [OBSERVER]
  
  make (v: INTEGER) -- Constructor
    do
      value := v
      create subscribers
    ensure
      subscribers.is_empty
    end
    
  update (v: INTEGER)
    do
      value := v
      across subscribers as o do o.notify end
    ensure
      value = v
    end
    
feature {OBSERVER}
  register (o: OBSERVER)
    require
      not subscribers.has (o)
    do
      subscribers.add (o)
    ensure
      subscribers.has (o)
    end    
end
\end{lstlisting}%
&%
\begin{lstlisting}
class OBSERVER

  subject: SUBJECT  
  cache: INTEGER
  
  make (s: SUBJECT) -- Constructor
    do
      subject := s
      s.register (Current)
      cache := s.value
    ensure
      subject = s
    end
    
feature {SUBJECT}    
  notify
    do
      cache := subject.value
    ensure
      subject = old subject
      cache = subject.value
    end
  
invariant
  subject.subscribers.has (Current)
  cache = subject.value
end
\end{lstlisting}
\end{tabular}
\fi
\lstset{basicstyle=\footnotesize}
\vspace{-8mm}
\caption{The \emph{Observer pattern}: an observer's \e{invariant} depends on the state of the \e{SUBJECT}, which reports its state changes to all its \e{subscribers}.
The clients of the subscribers must be able to rely on their \e{cache} always being consistent, while oblivious of the update/notify mechanisms that preserve invariants.} 
\label{fig:observer}
\end{figure}

\fakepar{\emph{Observer} pattern}
\cref{fig:observer} shows the essential parts of an implementation of the Observer design pattern in Eiffel.
An arbitrary number of \e{OBSERVER} objects (called ``subscribers'') monitor the public state of a single instance of class \e{SUBJECT}.
Each subscriber maintains a copy of the subject's relevant state (integer attribute \e{value} in \cref{fig:observer}) into one of its local variables (attribute \e{cache} in \cref{fig:observer}).
The subscribers' copies are cached values that must be consistent with the state of the subject, formalized as the invariant clause \e{cache = subject.value} of class \e{OBSERVER}, which depends on another object's state.
This dependency is not adequately captured by ownership schemes,
since no one subscriber can have exclusive control over the subject.

In the Observer pattern, consistency is maintained by means of explicit collaboration: 
the subject has a list of \e{subscribers}, updated whenever a new subscriber registers itself by calling \e{register (Current)}%
\footnote{\e{Current} in Eiffel denotes the current object (\lstinline[language=Java]|this| in Java and C\#).}
on the subject.
Upon every change to its state (method \e{update}),
the subject takes care of explicitly notifying all registered subscribers 
(using an \e{across} loop that calls \e{notify} on every \e{o} in \e{subscribers}).
This explicit collaboration scheme---called ``considerate programming'' in~\cite{Considerate}---%
ensures that the subscribers' state remains consistent (i.e., the class invariant holds) between calls to the public methods of the object structure.

\ifextended 
\cref{fig:observer} uses Eiffel's \emph{selective exports}\footnote{Similar to friend classes in C++.}
to separate the public interface of the classes from the methods internal to the object structure: 
\e{feature \{OBSERVER\}} denotes that method \e{register} is only available to instances of class \e{OBSERVER}, 
and \e{feature \{SUBJECT\}} similarly limits the visibility of \e{notify} to the subject.
While selective exports help emphasize collaboration patterns, they are not necessary for the discussion of the present paper, 
whose results are applicable to any object-oriented language regardless of the available visibility specifiers.
\fi

A methodology to verify the Observer pattern must ensure invariant stability; 
namely, that clients of \e{OBSERVER} can rely on its invariant without knowledge of the register/notify mechanism.
Another challenge is dealing with the fact that the number of subscribers attached to the subject is not fixed a priori, 
and hence we cannot produce explicit syntactic enumerations of the subscribers' \e{cache} attributes.
We must also be able to verify \e{update} and \e{notify} without relying on the class invariant as precondition---in fact, those methods are called on inconsistent objects precisely to restore consistency.

\begin{figure}[!tb]
\lstset{basicstyle=\scriptsize,xleftmargin=8mm}
\centering
\ifextended 
\begin{tabular}{p{.49\textwidth} p{.49\textwidth}}
\begin{lstlisting}
class COLLECTION [G]

  count: INTEGER
  
  make (capacity: INTEGER) -- Constructor
    require
      capacity >= 0      
    do
      create elements(1, capacity)
    ensure
      elements.count = capacity
      count = 0
    end

  remove_last
    require
      count > 0
    do
      count := count - 1
    ensure
      count = old count - 1
    end
  
feature {ITERATOR}
  elements: ARRAY [G]
  
invariant
  0 <= count and count <= elements.count
end

\end{lstlisting}%
&%
\begin{lstlisting}
class ITERATOR [G]

  target: COLLECTION [G]  
  before, after: BOOLEAN
        
  make (t: COLLECTION) -- Constructor
    do
      target := t ; before := True
    ensure
      target = t
      before and not after
    end
    
  item: G
    require
      not (before or after)
    do
      Result := target.elements [index]
    end    
  
feature {NONE}
  index: INTEGER
  
invariant
  0 <= index and index <= target.count + 1
  before = index < 1
  after = index > target.count
end

\end{lstlisting}
\end{tabular}
\fi
\lstset{basicstyle=\footnotesize}
\vspace{-8mm}
\caption{The \emph{Iterator pattern}: an iterator's invariant depends on the state of the collection it traverses, which is oblivious of the iterators.
Verification must prove that clients do not access disabled iterators, without knowing collection's and iterator's internal states.}
\label{fig:iterator}
\end{figure}

In the \textbf{\emph{Iterator} pattern}, an arbitrary number of iterator objects traverse a collection of elements.
\cref{fig:iterator} sketches an implementation where the \e{COLLECTION} uses an \e{ARRAY} of \e{elements} as underlying representation.
The \e{ITERATOR}'s main capability is to return the \e{item} at the current position \e{index} in the \e{target} collection%
\footnote{We omit the description of other necessary operations, such as advancing the iterator, since they are irrelevant for our discussion about invariants.}.
\e{item}'s precondition (\e{require}) specifies that this is possible only when the iterator points to a valid element of \e{target}, that is \e{index} is between \e{1} and \e{target.count} (included);
otherwise, if \e{index} is \e{0} the iterator is \e{before} the list, and if it equals \e{target.count + 1} it is \e{after} the list. 
The invariant of class \e{ITERATOR} defines the public state components \e{before} and \e{after} in terms of the internal state component \e{index}, 
as well as the acceptable variability range for \e{index}.

Since the iterator's invariant depends on the state of the target collection, modifying the collection (for example, by calling \e{remove_last}) may \emph{disable} the iterator (make it inconsistent).
This is aligned with the intended usage of iterators, which should be discarded after traversing a collection without changing it.
A verification methodology should ensure that clients of \e{ITERATOR} only access iterators in a consistent state, without knowledge of the iterator's internal state \e{index} or of its relation to the \e{target} collection.
\ifextended 
In fact, the selective exports used in \cref{fig:iterator} hide the details of \e{ITERATOR}'s invariant from its clients (the visibility of an invariant clause is determined by its least visible subexpression, and \e{feature \{NONE\}} denotes purely private members).
\else
\fi
An additional obstacle to verification comes from the fact that considerate programming would be at odds with the ephemeral nature of iterators compared to observers: 
collections are normally implemented unaware of the iterators operating on them; 
a flexible invariant methodology should allow such implementations.

\section{Existing Approaches}
\label{sec:existing}

\ifextended
This section reviews the main existing methodologies for specifying and reasoning about class invariants;  based on their most important features and limitations. 
\cref{sec:methodology} will present our own methodology.
For lack of space, we only discuss methodologies for inter-object dependencies that support modular reasoning (where local checks on individual classes or small groups of classes subsume global program correctness).
\fi

A crucial issue is deciding \emph{when} (at which program points) class invariants should hold: state-changing operations normally consist of sequences of elementary updates, which individually may break the class invariant temporarily.
To deal with this problem, some methodologies restrict the program points where class invariants are expected to hold; others interpret the invariants in a weakened form, which holds vacuously at intermediate steps during updates (and fully at crucial points).

\ifextended
Methodologies based on \textbf{visible-state semantics} only require invariants to hold when no operation is being executed on their objects, that is in states visible to clients.
This idea was introduced for Eiffel~\cite{meyer:object-oriented:2000}, and later also adopted by JML~\cite{JML99}.
Without additional mechanisms, visible-state semantics can't achieve modularity in the presence of callbacks 
(the client making the callback is unaware of ongoing operations that may affect the invariant) 
and of inter-object dependencies 
(if $o_1$'s invariant depends on $o_2$, the former is also affected by operations on $o_2$ invisible to clients of $o_1$).
Existing solutions adopt aliasing control measures~\cite{Muller06} to deal with hierarchical object structures described by ownership.
Other solutions~\cite{Middelkoop06,Middelkoop08,Considerate}, for collaborative invariants, explicitly indicate which objects might be inconsistent at method call boundaries; 
for example, method \e{register (o: OBSERVER)} of class \e{SUBJECT} in \cref{fig:observer} would be annotated with \vb{broken o} to specify that argument \e{o}'s invariant may not hold when executing \e{register}.
These two families of solutions---for hierarchical and for collaborative object structures---based on visible-state semantics are not easily combined; this is a practical limitation, since many object-oriented systems consist of an interplay between both types of structure.
For example, continuing with \cref{fig:observer}, objects of class \e{SUBJECT} collaborate with \e{OBSERVER} objects but also own a \e{subscribers} list as part of their representation.
Thus, when reasoning about method \e{register}, we should be able to deal with the call \e{subscribers.add (o)} whose argument \e{o} is inconsistent (and hence \e{add} cannot assume \e{o}'s invariant); 
however, annotating \e{LIST}'s \e{add} by declaring its argument \vb{broken} goes against modularity, as class \e{LIST} should not need to know how and where it is used.
The difficulty of integrating hierarchical and collaborative models is the main limitation of visible-state methodologies, and likely a reason why, to our knowledge, they have not been implemented in any program verifier.
\fi

Another family of methodologies, collectively known as \textbf{Boogie methodologies} after the program verifier where they have originally been implemented, follow the approach of weakening the default semantics of invariants
so that they can be evaluated only when appropriate.
In a nutshell, all classes include a ghost Boolean attribute \vb{closed},\footnote{We follow VCC's terminology~\cite{VCC} whenever applicable; other works may use different names.} 
which denotes whether an object is in a consistent state; an invariant \e{inv} is then interpreted as the weaker \vb{closed $\Implies$ inv}, which vacuously holds for open (i.e., not closed) objects.
Methods explicitly indicate whether they expect relevant objects to be closed or open;
\ifextended
this approach is more conducive to modularity than visible-state semantics:
it does not impose consistency by default at method call boundaries and thus does not require methods to list \emph{all} possibly inconsistent objects in the entire program.
\fi

The original Boogie methodologies, implemented in the Spec\# system~\cite{SpecSharp}, are main\-ly based on \emph{syntactic} mechanisms to express ownership relations.
For example, following \cite{Barnett04}, we would annotate attribute \e{elements} of class \e{COLLECTION} in \cref{fig:iterator} with \vb{rep}, to denote that it belongs to \e{COLLECTION}'s internal representation; 
thus, modifying \e{elements} is only possible if the \e{COLLECTION} object owning it has been opened---a situation where \vb{closed $\Implies$ count <= elements.count} vacuously holds.
This solution only supports representations based on bounded sets of objects known a priori and directly accessible through attributes.
Follow-up work~\cite{Dynamic} partially relaxes these restriction 
introducing a form of quantification predicating over an \vb{owner} ghost attribute (which goes up the ownership hierarchy), and a mechanism to transfer ownership.
\ifextended
The additional expressiveness comes with a price to pay mainly in terms of complex invariant admissibility conditions (hence, it may be hard to understand what is expressible and how) and complicated soundness proofs of the methodology.
\fi

In contrast, the VCC verifier~\cite{VCC} implements a Boogie methodology where ownership is encoded on top of LCI's \emph{semantic} approach~\cite{Cohen10}.
Objects include an additional ghost attribute, \vb{owns}, storing the set of all owned objects; ghost code modifies this set explicitly when the owner object is open.
In the example of \cref{fig:iterator}, instead of annotating attribute \e{elements} with \vb{rep}, 
we would introduce a first-order formula, such as \vb{owns = \{elements\}}, in the invariant of \e{COLLECTION} to express that \e{elements} is part of the representation.
The advantage of this approach becomes apparent with linked structures where owned elements are accessible only by following chains of references (e.g., a linked list owns all reachable cells).
In fact, semantic approaches to ownership provide the flexibility necessary to specify an unbounded number of owned objects, which may even be not directly attached to the owner, as well as to implement ownership transfers without need for ad hoc mechanisms.
They also simplify the rules of reasoning; 
for example, invariant admissibility becomes a simple proof obligation that 
all objects whose state is mentioned in the invariant are bound, by the same invariant, to belong to \vb{owns}.
These features have contributed to making VCC applicable to real-world systems~\cite{HyperV}.

In addition to ownership, some Boogie methodologies also deal with collaborating objects.
\cite{Dynamic} introduces the notion of \emph{visibility-based} invariants,
which requires that a class be aware of the types and invariants of all objects concerned with its state%
\footnote{We say that an object $o$ is \emph{concerned} with an attribute $a$ of another object $s$ if updating $s.a$ might affect $o$'s invariant.}.
For example, in \cref{fig:observer} \e{SUBJECT} must declare its \e{value} attribute with a modifier \e{dependent OBSERVER}.
Whenever the subject changes its \e{value}, it has to check that all potentially affected \e{OBSERVER}s are open.
If aware of the \e{OBSERVER}'s invariant, it can show that the only affected observers are $\{o\e{: OBSERVER} \mid o\e{.subject} = \e{Current}\}$.
Such indirect representations of the concerned objects complicate discharging the corresponding proof obligations;
and relying on knowing the concerned objects' invariants introduces tight coupling between the collaborating classes.
To lift these complications,
\cite{Friends} suggests instead to introduce a ghost attribute \e{deps} storing the set of all concerned objects.
It also introduces \emph{update guards}, allowing a concerned object to state conditions under which its invariant is preserved without revealing the invariant itself.
Both approaches \cite{Dynamic,Friends} have shortcomings that derive from their reliance on syntactic mechanisms and conditions: collaboration invariants can only depend on a bounded number of objects known a priori and accessible through attributes (called ``pivot fields'' in \cite{Friends}); 
the types of the concerned objects must be known explicitly;
and the numerous ad hoc annotations (e.g., \vb{friend} and \vb{keeping}) and operations (e.g., to modify \vb{deps}) make the methodologies harder to present and use.
One of the main goals of our methodology (\cref{sec:methodology}) is to lift these shortcomings by dealing with collaborative invariants by \emph{semantic} rather than syntactic means---similarly to what VCC did to the classic syntactic treatment of ownership.
\ifextended
The semantic approach makes SC very flexible, capable of accommodating disparate object-oriented design patterns without requiring ad hoc mechanisms.
\fi

Somewhat orthogonally to other Boogie-family approaches, 
the \emph{history invariants} methodology~\cite{History} provides for more loose coupling between the collaborating classes,
but gives up stability of invariants.

\section{Semantic Collaboration}
\label{sec:methodology}

Our novel invariant methodology belongs to the Boogie family; as we illustrated in \cref{sec:existing}, this entails that objects can be \emph{open} or \emph{closed}, and class invariants have to hold only for closed objects.
On top of semantic mechanisms for ownership, similar to those developed for VCC (see \cref{sec:existing}), our methodology also provides a semantic treatment of dependencies among collaborating objects; hence its name \emph{semantic collaboration}.
The keywords and constructs specific to SC are \texttt{\underline{underlined}} in the following.

\fakepar{Overview of semantic collaboration}
To specify collaboration patterns, we equip every object \e{o} with ghost fields \e{subjects} and \e{observers}.
As their names suggest,%
\footnote{While the names are inspired by the Observer pattern, they are also applicable to other collaboration patterns, as we demonstrate in \cref{sec:methodology:examples}. 
The formatting should avoid confusion.}
\e{o.subjects} stores the set of objects on which \e{o}'s invariant might depend; 
and \e{o.observers} stores the set of objects potentially concerned with \e{o} (analogous to \e{deps} in~\cite{Friends}).
The methodology achieves modularity by reducing global validity (all closed objects satisfy their invariants) to local checks of two kinds: 
\begin{inparaenum}[(\itshape i)]
\item \label{overview:admiss} all concerned objects are stored in \e{observers}; and 
\item \label{overview:updates} updates to the attributes of an object \e{o} maintain the validity of \e{o} and its observers.
\end{inparaenum}
Check~{(\itshape \ref{overview:admiss})} becomes an admissibility condition that every declared class invariant must satisfy.
Check~{(\itshape \ref{overview:updates})} holds vacuously for for open observers,
thus one way to satisfy it is to ``notify'' all observers of a potentially destructive update by opening them. 
For more flexibility the methodology also allows subjects to skip ``notifying'' observers whenever the attribute update satisfies its \emph{guard} (a notion also inspired by~\cite{Friends}).
This option is supported by another admissibility condition: 
an invariant must remain valid after updates to subjects that comply with their update guards.

\subsection{Preliminaries and Definitions}
\label{sec:prel-defin}

\ifextended
As it is customary, the following presentation targets fundamental constructs, 
while ignoring those that do not affect reasoning about invariants (e.g., control structures).
We also largely ignore issues related to inheritance, but we briefly come back to them in \cref{sec:conclusions}.
\fi

A program is a collection of classes.
A class is a collection of attributes, methods, and logical functions (side-effect free and terminating).
\ifextended
Any of those constructs can be declared \emph{ghost} if it is meant to be used only in specifications.
\fi

\fakepar{Built-in attributes}
Every class is implicitly equipped with ghost attributes: \e{closed} (to encode consistency); \e{owns} and \e{owner} (to encode the ownership hierarchy); and \e{subjects} and \e{observers} (to encode collaboration).
We also define the shorthands: \e{o.open} for \e{$\lnot$ o.closed}; \e{o.free} for \e{o.owner.open}; and \e{o.wrapped} for \e{o.closed $\,\land\,$ o.free}.
The \emph{ownership domain} of an object \e{o} is $\{\e{o}\}$ if \e{o} is open, and the transitive closure of \e{o.owns} if \e{o} is closed. 
Attributes \e{closed} and \e{owner} are only changed indirectly through the implicitly defined ghost methods \e{wrap} and \e{unwrap}, whose semantics is defined below.

\fakepar{Specifications}
The specification of a \emph{logical function} consists of a \emph{definition} (a side-effect free expression defining the function value) and a \e{read} clause (an expression that denotes the set of objects on which the value of the function may depend).
The specification of a \emph{method} consists of a \e{require} clause (a precondition), an \e{ensure} clause (a postcondition), and a \e{modify} clause (an expression that denotes the set of objects that the method may modify).
The specification of a \emph{class} includes its invariant \e{inv}.
The specification of an \emph{attribute} \e{a} consists of an \emph{update} \e{guard} (a Boolean expression over \e{Current} object, new attribute value \e{y}, and generic observer object \e{o}---written $\guard{\e{Current.a := y}, \e{o}}$).

\fakepar{Expressions}
In addition to the standard programming-language expressions, we support a restricted form of quantification through the syntax \e{all x $\,\in\,$ s : B(x)} for universal and 
\e{some x $\,\in\,$ s : B(x)} for existential quantification, where \e{s} is a set expression and \e{B(x)} is a Boolean expression over \e{x}.
The special expression \e{Void} (analogous to \lstinline[language=Java]|null| in Java and C\#) denotes an object that is always allocated and open.

The \emph{read set} $\reads{e}$ of a primitive expression $e$ is defined as follows: 
for an access \e{x.a} to attribute \e{a}, $\reads{\e{x.a}} = \{\e{x}\}$; 
for a call \e{x.f (y)} to logical function \e{f}, $\reads{\e{x.f (y)}}$ is given by the \e{f}'s \e{read} clause.
The read set of a compound expression $e$ is the union of the read sets of $e$'s subexpressions.

The current \emph{heap} $\Heap$ in which expressions are evaluated is normally clear from the context and left implicit.
Otherwise, $e_h$ denotes the value of expression $e$ in heap $h$; and $h[\e{x.f} \mapsto e]$ denotes the heap that agrees with $h$ everywhere except possibly about the value of \e{x.f}, which is $e$.
\ifextended 
Since we ignore deallocation, our heaps have no dangling references: only allocated objects are reachable from allocated objects.
\fi

\fakepar{Instructions}
For the present discussion, we only have to consider method calls \e{x.m (y)},
as well as \emph{heap update instructions}:
\e{create x} (allocate an object and attach it to \e{x}); 
\e{x.a := y} (update attribute \e{a});
and \e{x.wrap} and \e{x.unwrap} (opening and closing an object).
\ifextended

The \emph{write set} $\writes{s}$ of an primitive instruction $s$ is defined as follows: for an update \e{x.a := y} of attribute \e{a}, $\writes{\e{x.a := y}} = \{\e{x}\}$;
for opening or closing an object $x$, $\writes{\e{x.unwrap}} = \writes{\e{x.wrap}} = \{\e{x}\}\cup\e{x.owns}$;
for a call \e{x.r (y)} to method (or constructor) \e{r}, $\writes{\e{x.r (y)}}$ is the union of the ownership domains of all objects mentioned in \e{r}'s \e{modify} clause.
The write set of a compound instruction $s$ is the union of the write sets of the instructions in $s$.
\fi

\subsection{Semantic Collaboration: Goals and Proof Obligations}
\label{sec:methodology:formal}

The \textbf{goal} of any invariant methodology is to provide \emph{modular} proof obligations to establish \emph{global} validity:
the property that every object in the program is \emph{valid} at every program point.
Following SC's approach, an object is valid if satisfies its invariant when closed;
thus global validity is defined as:
\begin{equation}
\forall o : o.\e{closed} \Implies o.\e{inv}
\tag{G1}\label{eq:G1} \\
\end{equation}

Additionally, maintaining ownership-based invariants requires strengthening global validity with the property
that whenever a parent object $p$ is closed all its owned objects are closed (and their \e{owner} attributes point back to $p$):
\begin{equation}
\forall o, p : p.\e{closed} \And o \in p.\e{owns} \Implies o.\e{closed} \And o.\e{owner} = p \tag{G2}\label{eq:G2}
\end{equation}

\fakepar{Proof obligations}
The proof obligations specific to SC consist of two types of checks: 
\begin{inparaenum}[(\itshape i)]
\item \label{rules:admiss} every class invariant is \emph{admissible} according to \cref{def:admissibility}; and 
\item \label{rules:updates} every heap update instruction satisfies its precondition.
\end{inparaenum}
\ifextended
These proof obligations are \emph{modular} in that they only mention the state of the current object, its observers and owned objects.
\cref{sec:methodology:soundness} describes how establishing the local proof obligations entails global validity, that is subsumes checking \eqref{eq:G1} and \eqref{eq:G2}.
\fi

\ifextended
Admissibility captures the requirements that class invariants respect ownership and collaboration relations, modeled through ghost attributes \e{owns}, \e{subjects}, and \e{observers}.
\fi
\begin{Definition} \label{def:admissibility}
An invariant \e{inv} is \emph{admissible} iff:
\begin{enumerate}
\item \e{inv} only depends on \e{Current}, its owned objects, and its subjects:
\[
\e{inv} \quad \Implies\quad \reads{\e{inv}} \subseteq \left(\{\e{Current}\} \cup \e{owns} \cup \e{subjects}\right)
\tag{A1}\label{eq:I1}
\]

\item All subjects of \e{Current} are aware of it as an observer:
\[
\e{inv} \quad \Implies\quad \forall s : s \in \e{subjects} \Implies \e{Current} \in s.\e{observers}
\tag{A2}\label{eq:I2}
\]

\item \e{inv} is preserved by any update \e{s.a := y} that conforms to its guard:
\[
\forall s, a, y : s \in \e{subjects} \And \e{inv} \And \guard{s.a\:\e{:=}\:y, \e{Current}} \Implies \e{inv}_{\Heap[s.a \mapsto y]}
\tag{A3}\label{eq:I3}
\]

\item (Syntactic check) \e{inv} does not mention attributes \e{closed} and \e{owner}, directly or as part of the definitions of the mentioned logical functions.
\end{enumerate}
\end{Definition}

The specifications of the heap update instructions are given below;
the instructions only modify objects and attributes mentioned in the postconditions.

\begin{description}
\setlength{\tabcolsep}{15pt}
\item[Allocation] creates an open object owned by \e{Void} (and thus free), with no observers:\\
\begin{tabular}{lll}
\e{create x}  &  \e{require}  & \e{ensure} \\
              & \e{True}      & $\e{x.open}\And\e{x.owner = Void}\And\e{x.observers = \{\}}$ \\
\end{tabular}

\item[Unwrapping] opens a wrapped object: \\
\begin{tabular}{lll}
\e{x.unwrap}    &  \e{require}   & \e{ensure} \\
                &  \e{x.wrapped} & \e{x.open}
\end{tabular}

\setlength{\tabcolsep}{10pt}
\item[Attribute update] operates on an open object and preserves validity of its observers: \\
\begin{tabular}{lll}
\e{x.a := y}      & \e{require} & \e{ensure} \\
(\e{a /= closed}) & \e{x.open}  & \e{x.a = y}  \\
                  & $\e{all o} \in \e{x.observers} : \e{o.open} \Or \guard{\e{x.a := y}, \e{o}}$
\end{tabular}

\setlength{\tabcolsep}{15pt}
\item[Wrapping] closes an open object, whose invariant holds, and gives it ownership over all objects in its \e{owns} set: \\
\begin{tabular}{lll}
\e{x.wrap}  & \e{require}                                   & \e{ensure}      \\
            & $\e{x.open}\And\e{x.inv}$                     & \e{x.wrapped} \\
            & $\e{all o} \in \e{x.owns} : \e{o.wrapped}$    & $\e{all o} \in \e{x.owns} : \e{o.owner} = \e{x}$ \\
\end{tabular}
\end{description}

\ifextended
\fakepar{Other proof obligations}
The other proof obligations, which do not involve invariants, are the usual ones of axiomatic reasoning: every call to a method \e{m} occurs in a state that satisfies \e{m}'s precondition; 
executing a method \e{m} in a state that satisfies its precondition leads to a state that satisfies \e{m}'s postcondition; 
the \e{read} clause of every logical function \e{f} is consistent (i.e., the read set of \e{f}'s definition is a subset of \e{f}'s \e{read} clause); 
the \e{modify} clause of every method \e{m} is consistent (i.e., the write set of \e{m}'s body is a subset of \e{m}'s \e{modify} clause); 
and the definitions of logical functions are terminating.
\fi

\subsection{Soundness Argument}
\label{sec:methodology:soundness}

The soundness argument has to establish that every program that satisfies the proof obligations of SC is always globally valid, that is satisfies \eqref{eq:G1} and \eqref{eq:G2}.
We outline a proof of this fact in three parts.

The first part concerns ownership: every methodology that, like SC, imposes a suitable discipline of wrapping and unwrapping to manage ownership domains reduces \eqref{eq:G2} to local checks.

\begin{lemma}
\label{lemma:ownership}
Consider a methodology $M$ whose proof obligations verify the following:
\begin{enumerate}[a.]
\item \label{own:r1} freshly allocated objects are \e{open};
\item \label{own:r2} whenever \e{x.owner} is updated or \e{x.closed} is set to \e{False}, object \e{x} is \e{free};
\item \label{own:r3} whenever \e{x.closed} is updated to \e{True}, every object \e{o} in \e{x.owns} is \e{closed} and satisfies \e{o.owner = x};
\item \label{own:r4} whenever an attribute \e{x.a} (with $\e{a} \notin \{\e{closed}, \e{owner}\}$) is updated, object \e{x} is \e{open}.
\end{enumerate}
Then every program that satisfies $M$'s proof obligations also satisfies \eqref{eq:G2} everywhere.
\end{lemma}
\ifextended
\begin{proof}
The proof is by induction on the length of program traces.

The base case is the trace only consisting of the initial heap where no object is allocated but for an open object \e{Void}; thus \eqref{eq:G2} holds initially.
For the inductive step, let $h$ be the final heap of a trace where \eqref{eq:G2} invariably holds.
Consider an instruction $s$ that yields heap $h'$ if executed on $h$.
Without loss of generality, let $h' \neq h$; therefore, $s$ is either an allocation of a new object or an attribute update.
If $s$ allocates a new object \e{x}, \eqref{eq:G2} still holds in $h'$: \e{x} is open (rule~\textit{\ref{own:r1}}) and is in no other object's \e{owns} set, since \e{x} has just been created.
If $s$ is an attribute update, it can only invalidate \eqref{eq:G2} if it updates \e{closed}, \e{owns}, or \e{owner}.
If $s$ updates some $o$\e{.owner} in \eqref{eq:G2}'s consequent or sets $o$\e{.closed} to \e{False}, then $o$ is \e{free} (rule~\textit{\ref{own:r2}}); thus $o$\e{.owner} is open, and hence \eqref{eq:G2}'s antecedent is false.
If $s$ sets to \e{True} some $p$\e{.closed} in \eqref{eq:G2}'s antecedent, then rule~\textit{\ref{own:r3}} implies the whole \eqref{eq:G2} holds.
If $s$ updates some $p$\e{.owns} in \eqref{eq:G2}'s antecedent, then $p$ is open (rule~\textit{\ref{own:r4}}); thus, \eqref{eq:G2}'s antecedent is false.
\qed
\end{proof}
\fi

The second part applies to any kind of inter-object invariants and assumes a methodology that, like SC, checks that attribute updates preserve validity of all concerned objects; 
we show that such checks subsume \eqref{eq:G1}.
How a methodology identifies concerned objects is left unspecified as yet.
\begin{lemma}
\label{lemma:soundness_general}
Consider a methodology $M$ whose proof obligations verify the following:
\begin{enumerate}[a.]
\item \label{sg:r1} freshly allocated objects are \e{open};
\item \label{sg:r2} whenever \e{x.closed} is updated to \e{True}, \e{x.inv} holds;
\item \label{sg:r3} whenever an attribute \e{x.a} (with \e{a} $\neq$ \e{closed}) is updated to some \e{y}, every concerned object satisfies $(\e{o.closed} \And \e{o.inv}) \Implies \e{o.inv}_{\Heap[\e{x.a} \mapsto \e{y}]}$;
\item \label{sg:r4} class invariants depend neither on attribute \e{closed} nor on the allocation status of objects.
\end{enumerate}
Then every program that satisfies $M$'s proof obligations also satisfies \eqref{eq:G1} everywhere.
\end{lemma}
\ifextended
\begin{proof}
The proof is by induction on the length of program traces.

The base case is the trace only consisting of the initial heap where no object is allocated but for an open object \e{Void}; thus \eqref{eq:G1} holds initially.
For the inductive step, let $h$ be the final heap of a trace where \eqref{eq:G1} invariably holds.
Consider an instruction $s$ that yields heap $h'$ if executed on $h$.
Without loss of generality, let $h' \neq h$; therefore, $s$ is either an allocation of a new object or an attribute update.
If $s$ allocates a new object \e{x}, \eqref{eq:G1} still holds in $h'$: \e{x} is open (rule~\textit{\ref{sg:r1}}) and no other object's invariants depends on it, since \e{x} has just been created and class invariant do not know about allocation status (rule~\textit{\ref{sg:r4}}).
If $s$ sets to \e{False} some $o$\e{.closed} in \eqref{eq:G1}'s antecedent, then \eqref{eq:G1} vacuously hold.
If $s$ sets to \e{True} some $o$\e{.closed} in \eqref{eq:G1}'s antecedent, then $o$\e{.inv} holds (rule~\textit{\ref{sg:r2}}); thus \eqref{eq:G1} holds too.
Also, updates to some $o$\e{.closed} cannot concern the invariants of objects other than $o$ (rule~\textit{\ref{sg:r4}}).
If $s$ updates some \e{x.a}, with $\e{a} \neq \e{closed}$, let $o$ be any object concerned with the update;
either $o$ is open, or it is closed and $o$\e{.inv} holds in $h$ by the induction hypothesis, so rule~\textit{\ref{sg:r3}} applies.
Either way, \eqref{eq:G1} holds in $h'$ for $o$.
\qed
\end{proof}
\fi

The third part of the soundness proof argues that SC satisfies the hypotheses of \cref{lemma:ownership,lemma:soundness_general}, and hence ensures global validity.
\begin{proposition}
Every program that satisfies the proof obligations of SC also satisfies \eqref{eq:G2} and \eqref{eq:G1} everywhere.
\end{proposition}
\ifextended
\begin{proof}
SC satisfies the hypotheses of \cref{lemma:ownership}: allocation satisfies rule~\textit{\ref{own:r1}}; 
unwrapping satisfies rule~\textit{\ref{own:r2}} and wrapping satisfies rules~\textit{\ref{own:r2}} and~\textit{\ref{own:r3}}
(we assume that \e{wrap} first updates the \e{owner} attribute of every object in the \e{owns} set of its argument, and then updates the \e{closed} attribute of its argument); 
remember that \e{closed} and \e{owner} are only changed by wrap and unwrap. 
Attribute update satisfies rule~\textit{\ref{own:r4}}.

It also satisfies the hypotheses of \cref{lemma:soundness_general}: allocation satisfies rule~\textit{\ref{sg:r1}}; wrapping satisfies rule~\textit{\ref{sg:r2}}; invariant admissibility and the rules of language syntax satisfy rule~\textit{\ref{sg:r4}}.
Rule~\textit{\ref{sg:r3}} requires more details.
First note that invariant admissibility requires that no invariant mention \e{owner}; thus no object is concerned with wrapping (the only operation that can change \e{owner}), which therefore vacuously satisfies rule~\textit{\ref{sg:r3}}.
Now, consider an update \e{x.a := y} with $\e{a} \neq \e{owner}$ and $\e{a} \neq \e{closed}$, and let \e{o} be any concerned object.
Assuming \e{o.closed} and \e{o.inv} hold for a generic heap $h$, we have to show that \e{o.inv} also holds of the heap $h' = h[\e{x.a} \mapsto \e{y}]$.
By definition of read set, $\e{x} \in \reads{\e{o.inv}}$; \e{o.inv} is also admissible and hence it satisfies \eqref{eq:I1}; therefore $\e{x} \in \{\e{o}\} \cup \e{o.owns} \cup \e{o.subjects}$.
However, the first precondition of the attribute update rule says that \e{x} is open; thus $\e{x} \neq \e{o}$ because \e{o} is closed.
We already proved that $h$ satisfies \eqref{eq:G2}; for $p = \e{o}$ this entails that all objects in \e{o.owns} are closed; therefore, $\e{x} \not\in \e{o.owns}$ as well.
We conclude that $\e{x} \in \e{o.subjects}$ which, combined with condition \eqref{eq:I2} for \e{o.inv}'s admissibility, implies that $\e{o} \in \e{x.observers}$ holds in $h$.
Finally, the second precondition of the attribute update rule establishes $\guard{\e{x.a := y}, \e{o}}$,
and thus by admissibility condition \eqref{eq:I3}, \e{o.inv} still holds in in the heap $h'$.
\qed
\end{proof}
\fi

\ifextended
As a closing remark, we note that another way to show soundness of SC is via reduction to LCI.
To encode collaboration in LCI on top of the ownership encoding detailed in~\cite{Cohen10},
we add the following clauses to the invariant of each class:
one stating that all \e{subjects} know \e{Current} for an observer (the consequent of \eqref{eq:I2}),
and for each attribute of \e{Current}, another one stating that all \e{observers} approve of the changes to this attribute.
\fi

\subsection{Examples}
\label{sec:methodology:examples}

We illustrate SC on the two examples of \cref{sec:examples}:
\cref{fig:observer_semcol,fig:iterator_semcol} show the Observer and Iterator patterns fully annotated according to the rules of \cref{sec:methodology:formal}.
We use the shorthands \e{wrap_all (s)} and \e{unwrap_all (s)} to denote calls to \e{wrap} and \e{unwrap} on all objects in a set \e{s}.
As we discuss in \cref{sec:evaluation}, several annotations of \cref{fig:observer_semcol,fig:iterator_semcol} are subsumed by the defaults mentioned in \cref{sec:methodology:defaults}.
We postpone to \cref{sec:methodology:guards} dealing with update guards and the corresponding admissibility condition \eqref{eq:I3}.

\begin{figure}[!tb]
\lstset{basicstyle=\scriptsize,xleftmargin=8mm}
\centering
\ifextended 
\begin{tabular}{p{.53\textwidth} p{.45\textwidth}}
\begin{lstlisting}
class SUBJECT

  value: INTEGER
  subscribers: LIST [OBSERVER]
  
  make (v: INTEGER) -- Constructor
    require open
    modify Current
    do      
      value := v
      create subscribers     
      owns := { subscribers }
      wrap 
    ensure
      subscribers.is_empty
      wrapped
    end
    
  update (v: INTEGER)
    require
      wrapped
      all o $\in$ observers : o.wrapped
    modify Current, observers
    do
      unwrap ; unwrap_all (observers)
      value := v
      across subscribers as o do o.notify end
      wrap_all (observers) ; wrap
    ensure
      value = v
      wrapped
      all o $\in$ observers : o.wrapped
      observers = old observers
    end
    
feature {OBSERVER}
  register (o: OBSERVER)
    require
      not subscribers.has (o)
      wrapped
      o.open
    modify Current
    do
      unwrap
      subscribers.add (o)
      observers := observers + { o }
      wrap
    ensure
      subscribers.has (o)
      wrapped
    end    
\end{lstlisting}%
&%
\begin{lstlisting}
invariant
  observers = subscribers.range
  owns = { subscribers }
  subjects = {}
end    

class OBSERVER

  subject: SUBJECT  
  cache: INTEGER
  
  make (s: SUBJECT) -- Constructor
    require
      open
      s.wrapped
    modify Current, s
    do
      subject := s
      s.register (Current)
      cache := s.value
      subjects := { s }
      wrap       
    ensure
      subject = s
      wrapped
      s.wrapped
    end
    
feature {SUBJECT}    
  notify
    require 
      open
      subjects = {subject}
      subject.observers.has (Current)
      observers = {}
      onws = {}
    modify Current
    do
      cache := subject.value
    ensure
      inv
    end
  
invariant
  cache = subject.value
  subjects = { subject }
  subject.observers.has (Current)
  observers = {}
  owns = {}
end
\end{lstlisting}
\end{tabular}
\fi
\lstset{basicstyle=\footnotesize}
\vspace{-8mm}
\caption{The \emph{Observer pattern} using SC annotations (underlined).}
\label{fig:observer_semcol}
\end{figure}

\fakepar{Observer pattern}
The \e{OBSERVER}'s invariant is admissible (\cref{def:admissibility}) because it ensures that \e{subject} is in \e{subjects} \eqref{eq:I1} and that \e{Current} is in the \e{subject}'s \e{observers} \eqref{eq:I2}.
Constructors normally wrap freshly allocated objects after setting up their state.
Public method \e{update} must be called when the whole object structure is wrapped and makes sure that it is wrapped again when the method terminates.
This specification style is convenient for public methods, as it allows clients to interact with the class while maintaining objects in a consistent state, without having to explicitly discharge any condition.
Methods such as \e{register} and \e{notify}, with restricted visibility, work instead with open objects and restore their invariants so that they can be wrapped upon return.
Since \e{notify} explicitly ensures \e{inv}, \e{update} does not need the precise definition of the observer's invariant in order to wrap it
(it only needs to know enough to establish the precondition of \e{notify}).
Thus the same style of specification would work if \e{OBSERVER} were an abstract class and its subclasses maintained different views of subject's \e{value}.

Let us illustrate the intuitive reason why an instance of \e{SUBJECT} cannot invalidate any object observing its state.
On the one hand, by the attribute update rule, any change to a subject's state (such as assignment to \e{value} in \e{update}) must be reconciled with its \e{observers}.
On the other hand, any closed concerned \e{OBSERVER} object must be contained in its \e{subject}'s \e{observers} set:
a subject cannot surreptitiously remove anything from this set,
since such a change would require an attribute update, and thus, again, would have to be reconciled with all current members of \e{observers}.

\ifextended
Note that we had to restate the first invariant clause of \e{OBSERVER} from \cref{fig:observer} in terms of \e{observers} instead of \e{subscribers}.
In general, collaboration invariants have to be expressed directly in terms of attributes of subjects and cannot refer to their ownership domains
(including through logical functions).
This is not a syntactic restriction but follows from the fact that it is rarely possible to establish a subject/observer relation with the whole domain
(in this example, we would have to require \e{LIST} to allow \e{OBSERVER} objects in its \e{observers} set).
This limitation can always be easily circumvented, however, by introducing a ghost attribute in the subject that mirrors the requires state.
\fi

\begin{figure}[!tb]
\lstset{basicstyle=\scriptsize,xleftmargin=8mm}
\centering
\begin{adjustwidth}{-4mm}{-4mm}
\ifextended 
\begin{tabular}{p{.45\textwidth} p{.6\textwidth}}
\begin{lstlisting}
class COLLECTION [G]

  count: INTEGER
  
  make (capacity: INTEGER) -- Constructor
    require
      open
      capacity >= 0      
    modify Current
    do
      create elements(1, capacity)
      owns := { elements } ; wrap 
    ensure
      elements.count = capacity
      count = 0
      observers = {}
    end

  remove_last
    require
      count > 0
      wrapped
      all o $\in$ observers : o.wrapped
    modify Current, observers
    do
      unwrap ; unwrap_all (observers)
      observers := {}
      count := count - 1
      wrap
    ensure
      count = old count - 1
      wrapped
      observers = {}
      all o $\in$ old observers : o.open
    end
  
feature {ITERATOR}
  elements: ARRAY [G]
  
invariant
  0 <= count and count <= elements.count
  owns = { elements }
  subjects = {}
end

\end{lstlisting}%
&%
\begin{lstlisting}
class ITERATOR [G]

  target: COLLECTION [G]  
  before, after: BOOLEAN
        
  make (t: COLLECTION) -- Constructor
    require
      open and t.wrapped
    modify Current, t
    do
      target := t
      before := True      
      t.unwrap
      t.observers := t.observers + { Current }
      t.wrap      
      subjects := { t }
      wrap
    ensure
      target = t
      before and not after
      wrapped
    end
    
  item: G
    require
      not (before or after)
      wrapped and t.wrapped
    do
      Result := target.elements [index]
    end    
  
feature {NONE}
  index: INTEGER
  
invariant
  0 <= index and index <= target.count + 1
  before = index < 1
  after = index > target.count
  subjects = { target }
  target.observers.has (Current)
  observers = {} and owns = {}
end

\end{lstlisting}
\end{tabular}
\fi
\end{adjustwidth}
\lstset{basicstyle=\footnotesize}
\vspace{-8mm}
\caption{The \emph{Iterator pattern} using SC annotations (underlined).}
\label{fig:iterator_semcol}
\end{figure}

\fakepar{Iterator pattern}
The main differences in the annotations of the Iterator pattern occur in the \e{COLLECTION} class whose non-ghost state is, unlike \e{SUBJECT} above, unaware of its \e{observers}.
Method \e{remove_last} has to unwrap its \e{observers} according to the update rule.
However, it has no way of restoring their invariants (in fact, a collection is in general unaware even of the \emph{types} of the iterators operating on it).
Therefore, it can only leave them in an inconsistent state and remove them from the \e{observers} set.
Public methods of \e{ITERATOR}, such as \e{item}, normally operate on wrapped objects, and hence in general cannot be called after some operations on the collection has disabled its iterators.
The only way out of this is if the client of collection and iterators can prove that a certain iterator object \e{i_x} was not in the modified collection's \e{observers}; this is possible if, for example, the client directly created \e{i_x}.
The fact that now clients are directly responsible for keeping track of the \e{observers} set is germane to the iterator domain: iterators are meant to be used locally by clients.

\subsection{Default Annotations}
\label{sec:methodology:defaults}

\ifextended
The annotation patterns shown in \cref{sec:methodology:examples} occur frequently in object-oriented programs.
To reduce the annotation burden in those cases, we suggest the following defaults.

\begin{description}
\item[Pre- and postconditions:] public procedures (methods not returning values) require and ensure that \e{Current}, its \e{observers}, and method arguments be \e{wrapped}.
\item[Modify clauses:] procedures modify \e{Current}; functions (methods returning values) modify nothing.
\item[Invariants:] Built-in ghost set attributes (such as \e{owns}) are invariably empty if they are not mentioned in the programmer-written invariant.
\item[Wrapping:] public procedures start by unwrapping \e{Current} and terminate after wrapping it back.
\item[Built-in set manipulation:] if a built-in ghost set attribute $s$ is only mentioned in an invariant clause of the form $s = \e{expr}$, then $s$ is considered \emph{implicit}; correspondingly, every \e{wrap} of objects enclosing $s$ will implicitly perform an assignment $s \e{:= expr}$.\footnote{This is inspired by the default ``static'' treatment of \e{owns} sets in VCC.}
\end{description}

These defaults encourage considerate programming:
unless explicitly specified otherwise, an object is always required to restore the consistency of its observers at the end of a public method.
This is a useful property, since the considerate paradigm promotes encapsulation and is convenient for the clients. 
Nevertheless, the defaults are only optional suggestions that can be overridden by providing explicit annotations; this ensures that they do not tarnish the flexibility and semantic nature of our methodology.
\fi

\subsection{Update guards}
\label{sec:methodology:guards}

Update guards are used to distribute the burden of reasoning about attribute updates between subjects and observers,
depending on the intended collaboration scheme.
At one extreme, if a $\guard{\e{x.a := y}, \e{o}}$ is identically \e{False}, the burden is entirely on the subject, which must check that all observers are open whenever \e{a} is updated; in contrast, the admissibility condition \eqref{eq:I3} holds vacuously for the observer \e{o}.
At the other extreme, if a guard is identically \e{True}, the burden is entirely on the observer,
which deals with \eqref{eq:I3} as a proof obligation that its invariant does not depend on \e{a}; in contrast, the subject \e{x} can update \e{a} without particular constraints.

Another recurring choice for a guard is $\e{inv(o)}\Implies\e{inv(o)}_{\Heap[\e{x.a} \mapsto \e{y}]}$.
For its flexibility, we chose this as the default guard of SC.
Just like \e{False}, this guard also does not burden the observer, but is more flexible at the other end: upon updating, the subject can establish that each observer is either open or its invariant is preserved.
The subject can rely on the latter condition if the observer's invariants are known, and ignore it otherwise.

When it comes to built-in ghost attributes, \e{owns} and \e{subjects} are guarded with \e{True}, since other objects are not supposed to depend on them,
while \e{observers} has a more interesting guard, namely $\guard{\e{x.observers := y}, \e{o}} = \e{o} \in \e{y}$.
This guard reflects the way this attribute is commonly used in collaboration invariants,
while leaving the subject with reasonable freedom to manipulate it; 
for example, adding new observers to the set \e{observers} without ``notifying'' the existing ones
(this is used, in particular, in the \e{register} method of \cref{fig:observer_semcol}).

\section{Experimental Evaluation}
\label{sec:evaluation}

We arranged a collection of representative challenge problems involving inter-object collaboration, and we specified and verified them using our SC methodology. 
This section presents the challenge problems~(\cref{sec:evaluation:challenge}), and discusses their solutions using SC~(\cref{sec:evaluation:results}), 
\ifextended
as well as other methodologies, in particular those described in \cref{sec:existing} (\cref{sec:comp-with-exist}).
\fi
See~\cite{SC} for full versions of problem descriptions, 
together with our solutions, and a web interface to the AutoProof verifier.

\subsection{Challenge Problems}
\label{sec:evaluation:challenge}

Beside using it directly to evaluate SC, the collection of challenge problems described in this section can be a benchmark for other invariant methodologies.
The benchmark consists of six examples of varying degree of difficulty, which capture the essence of various collaboration patterns often found in object-oriented software.
The emphasis is on non-hierarchical structures that maintain a global invariant.

We briefly present the six problems in roughly increasing order of difficulty in terms of the shape of references in the heap, state update patterns, and challenges posed to preserving encapsulation.
\ifextended 

\begin{wrapfigure}{l}{25mm}
\begin{center}
\vspace{-0.3cm}

\begin{tikzpicture}
  \node [object, label=above:{subject}] (subject) {};
  \node [object, above right=of subject, label=above:{observer}] (observer-above) {};
  \node [object, below right=of subject, label=below:{observer}] (observer-below) {};
  \path (subject) edge[bend left=10] (observer-above);
  \path (observer-above) edge[bend left=10] (subject);
  \path (subject) edge[bend left=10] (observer-below);
  \path (observer-below) edge[bend left=10] (subject);
\end{tikzpicture}

\vspace{-0.5cm}
\end{center}
\end{wrapfigure}
\textbf{Observer}~\cite{History,Parkinson07,Middelkoop08} (see also SAVCBS~'07~\cite{SAVCBS}, and \cref{sec:examples}).
The invariants of the observer objects depend on the state of the subject.
Verification must ensure that the subject reports all its state changes to all observers, so that their clients can always rely on a globally consistent state.
As \emph{additional challenge}: combination with ownership 
(the subject keeps references to its observers in a collection, which is a part of its representation).

\emph{Variants}: a simplified version where the number of observers is fixed (thus collections of observers are not needed); a more complex version with multiple observer classes related by inheritance, each class redefining class invariant and implementation of the \e{notify} method.

\begin{wrapfigure}{l}{25mm}
\begin{center}
\vspace{-0.3cm}

\begin{tikzpicture}
  \node [object, label=above:{collection}] (collection) {};
  \node [object, above right=of collection, label=above:{iterator}] (iterator-above) {};
  \node [object, below right=of collection, label=below:{iterator}] (iterator-below) {};
  \path (iterator-above) edge (collection);
  \path (iterator-below) edge (collection);
\end{tikzpicture}

\vspace{-0.5cm}
\end{center}
\end{wrapfigure}
\textbf{Iterator}~\cite{History} (see also SAVCBS~'06~\cite{SAVCBS}, and \cref{sec:examples}).
Unlike observers in the Observer pattern, the implementation of a collection is not aware of the iterators operating on it.
Specification must still be able to refer to the iterators attached to the collection while avoiding global reasoning.
As \emph{additional challenge}: we cannot rely on the implementation following considerate programming (where objects must be in consistent states at public call boundaries).

\emph{Variants}: a more complex version where iterators may modify the collection.
\fi

\begin{wrapfigure}{l}{25mm}
\begin{center}
\vspace{-0.3cm}

\begin{tikzpicture}
  \node [object, label=above:{master}] (master) {};
  \node [object, above right=of master, label=above:{slave}] (clock-above) {};
  \node [object, below right=of master, label=below:{slave}] (clock-below) {};
  \path (clock-above) edge (master);
  \path (clock-below) edge (master);
\end{tikzpicture}

\vspace{-0.5cm}
\end{center}
\end{wrapfigure}
\textbf{Master clock}~\cite{Friends,History}.
The time stored by a master clock can increase (public method \e{tick}) or be set to zero (public method \e{reset}).
The time stored locally by each slave clock must never exceed the master's but need not be perfectly synchronized.
Therefore, when the master is \e{reset} its slaves are disabled until they synchronize (similar to iterators); when the master increments the time its slaves remain in a consistent state without requiring synchronization.
\emph{Additional challenges}: \e{tick}'s frame does not include slaves; perform reasoning local to the master  with only partial knowledge of the slaves' invariants.

\emph{Variants}: a simplified version without \e{reset} (slaves cannot become inconsistent).

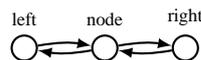
\begin{wrapfigure}{l}{30mm}
\begin{center}
\vspace{-0.3cm}

\begin{tikzpicture}
  \node [object, label=above:{node}] (node) {};
  \node [object, right=of node, label=above:{right}] (right) {};
  \node [object, left=of node, label=above:{left}] (left) {};
  \path (node) edge[bend left=10] (right);
  \path (right) edge[bend left=10] (node);
  \path (node) edge[bend left=10] (left);
  \path (left) edge[bend left=10] (node);
\end{tikzpicture}

\vspace{-0.3cm}
\end{center}
\end{wrapfigure}
\textbf{Doubly-linked list}~\cite{Dynamic,Middelkoop06}.
The specification expresses the consistency of the \e{left} and \e{right} neighbors directly attached to each \e{node}.
Verification establishes that updates local to a node (such as inserting or removing a node next to it) preserve consistency.
Unlike in the previous examples, the heap structure is recursive;
the main challenge is thus avoiding considering the list as a whole (such as to propagate the effects of local changes).

\begin{wrapfigure}{l}{35mm}
\begin{center}
\vspace{-0.3cm}

\begin{tikzpicture}
  [edge from parent/.style={latex-latex,thick,draw}]
\node [object] {}
  child {
    node [object] {}
      child { node [object] {} }
      child [missing]
  }
  child {
    node [object] {}
      child { node [object] {} }
      child { node [object] {} }
  };
\end{tikzpicture}

\vspace{-0.5cm}
\end{center}
\end{wrapfigure}
\textbf{Composite}~\cite{Need,Considerate,Leavens07}, (see also SAVCBS~'08~\cite{SAVCBS}).
A tree structure maintains consistency between the values stored by parent and children nodes (for example, the value of every node is the maximum of its children's).
Clients can add children anywhere in the tree; therefore, ownership is unsuitable to model this example.
Two new challenges are that the node invariant depends on an unbounded number of children; and that the effects of updates local to a node (such as adding a child) may propagate up the whole tree involving an unbounded number of nodes.
Specification deals with these unbounded-size footprints; and verification must also ensure that the propagation to restore global consistency terminates.
Clients of a tree can rely on a globally consistent state while ignoring the tree structure.

\emph{Variations}: a simplified version with $n$-ary trees for fixed $n$ (the number of children is bounded); 
more complex versions where one can also remove nodes or add \ifextended whole\fi subtrees.

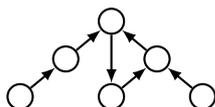
\begin{wrapfigure}{l}{35mm}
\begin{center}
\vspace{-0.3cm}

\begin{tikzpicture}
  [edge from parent/.style={latex-,thick,draw}]
\node [object] (root) {}
  child {
    node [object] {}
      child { node [object] {} }
      child [missing]
  }
  child {
    node [object] {}
      child { node [object] (bottom) {} }
      child { node [object] {} }
  };
  \path (root) edge (bottom);
\end{tikzpicture}

\vspace{-0.5cm}
\end{center}
\end{wrapfigure}
\textbf{PIP}~\cite{Need,Considerate}.
The Priority Inheritance Protocol~\cite{PIP} describes a compound whose nodes are more loosely related than in the Composite pattern: each node has a reference to at most one parent node, and cycles are possible.
Unlike in the Composite pattern, the invariant of a node depends on the state of objects not directly accessible in the heap (parents do not have references to their children).
New challenges derive from the possible presence of cycles,
and the need to add children that might already be connected to whole graphs;
specifying footprints and reasoning about termination \ifextended of update operations \fi are trickier. 

\subsection{Results and Discussion}
\label{sec:evaluation:results}

\begin{table}[!tb]
\caption{The challenge problems specified and verified using SC.}\label{tab:semcol_stats}
\begin{center}
\begin{tabular}{l|r|r|r|r|r|r|r|r}
        & \textsc{size}   & \multicolumn{4}{c|}{\textsc{tokens} (no defaults)} & \multicolumn{2}{c|}{\textsc{tokens} (with defaults)}          & \textsc{time}   \\
\cline{3-8}
\textsc{problem} & (LOC)  & \textsc{code}  & \textsc{req}  & \textsc{aux} & \textsc{spec/code}           & \textsc{aux} & \multicolumn{1}{c|}{\textsc{spec/code}}  & (sec.)               \\
\hline
Observer         & 129    & 156            & 52            & 296          & 2.2                          & 185          & 1.5                 & 8                   \\
Iterator         & 177    & 168            & 176           & 315          & 2.9                          & 247          & 2.5                 & 12                  \\
Master clock     & 130    & 85             & 69            & 267          & 4.0                          & 190          & 3.1                 & 6                   \\
DLL              & 147    & 136            & 83            & 435          & 3.8                          & 320          & 3.0                 & 18                  \\
Composite        & 188    & 124            & 270           & 543          & 6.6                          & 427          & 5.6                 & 18                  \\
PIP              & 152    & 116            & 310           & 445          & 6.5                          & 402          & 6.1                 & 18                  \\
\hline
\textbf{Total}   & 923    & 785            & 960           & 2301         & 4.2                          & 1771         & 3.5                 & 80                 \\
\end{tabular}
\end{center}
\end{table}

We specified the six challenge problems using SC, and verified the annotated Eiffel programs with AutoProof.
\cref{tab:semcol_stats} shows various metrics about our solutions: 
the \textsc{size} of each annotated program; 
the number of \textsc{tokens} of executable \textsc{code}, \textsc{req}uirements specification (the given functional specification to be verified), 
and \textsc{aux}iliary annotations (specific to our methodology, both with and without default annotations); 
the \textsc{spec/code} overhead, i.e., $(\textsc{req} + \textsc{aux})/\textsc{code}$; 
and the verification time in AutoProof.
The overhead is roughly between 1.5 (for Observer) and 6 (for PIP),
which is comparable with that of other verification methodologies applied to similar problems.
The default annotations of \cref{sec:methodology:defaults} reduce the overhead by a factor of 1.3 on average.

The PIP example is perfectly possible using ghost code, contrary to what is claimed elsewhere~\cite{Need}.
In our solution, every node includes a ghost set \e{children} with all the child nodes (inaccessible in the non-ghost heap); 
it is defined by the invariant clause \e{parent /= Void $\Implies$ parent.children.has (Current)}, which ensures that \e{children} contains every closed node $n$ such that $n\e{.parent} = \e{Current}$.
Based on this, the fundamental consistency property is that the \e{value} of each node is the maximum of the values of nodes in \e{children} (or a default value for nodes without children), assuming maximum is the required relation between parents and children.

The main challenge in Composite and PIP is reasoning about framing and termination of the state updates that propagate along the graph structure.
For framing specifications, we use a ghost set \e{ancestors} with all the nodes reachable following \e{parent} references.
Proving termination in PIP requires keeping track of all visited nodes and showing that the set of ancestors that haven't yet been visited is strictly shrinking.

\begin{table}[!tb]
\begin{center}
\begin{threeparttable}
\caption{Comparison of invariant protocols on the challenge problems.}\label{tab:comparison}
\begin{tabular}{l|c|c|c|c|c||c}
              & \multicolumn{2}{c|}{\textsc{visible-state semantics}}                                & \multicolumn{4}{c}{\textsc{Boogie methodologies}}\\
              \cline{2-7}
              & Cooperation~\cite{Middelkoop08} & Considerate~\cite{Considerate}  & Spec\#~\cite{Dynamic}   & Friends~\cite{Friends}  & History~\cite{History}  & SC\\
\hline
Observer      & \byes                           & \yes                            & \yes                    & \byes                   & \byes\tnote{d}          & \byes\\
Iterator      & \no\tnote{a}                    & \no\tnote{a}                    & \yes                    & \yes                    & \byes\tnote{d}          & \byes\\
Master clock  & \no\tnote{a}                    & \no\tnote{a}                    & \yes                    & \byes                   & \byes\tnote{d}          & \byes\\
DLL           & \yes                            & \yes                            & \byes                   & \yes                    & \yes\tnote{d}           & \byes\\
Composite     & \no\tnote{b}                    & \byes\tnote{c}                  & \no\tnote{b}            & \no\tnote{b}            & \no\tnote{b}            & \byes\\
PIP           & \no\tnote{b}                    & \byes\tnote{c}                  & \no\tnote{b}            & \no\tnote{b}            & \no\tnote{b}            & \byes\\ 
\end{tabular}
\begin{tablenotes}[para]
\item [a] Only considerate programming
\item [b] Only bounded set of reachable subjects
\item [c] No framing specification
\item [d] No invariant stability
\end{tablenotes}
\end{threeparttable}
\end{center}
\end{table}

\subsection{Comparison with Existing Approaches}
\label{sec:comp-with-exist}

\ifextended
We outline a comparison with existing approaches (focusing on those discussed in \cref{sec:existing}) on our six challenge problems.
\fi
\cref{tab:comparison} reports how each methodology fares on each challenge problem: \no{} for ``methodology not applicable'', \yes{} for ``applicable'', and \byes{} for ``applicable and used to demonstrate the methodology when introduced''.

Only SC is applicable to all the challenges, and other methodologies often have other limitations (notes in \cref{tab:comparison}).
Most approaches cannot deal with unbounded sets of subjects, and hence are inapplicable to Composite and PIP.
The methodology of~\cite{Considerate} is an exception as it allows set comprehensions in invariants; however, it lacks an implementation and does not discuss framing, which constitutes a major challenge in Composite and PIP.
Both methodologies~\cite{Middelkoop08,Considerate} based on visible-state semantics are inapplicable to implementations which do not follow considerate programming; 
they also lack support for hierarchical object dependencies, and thus cannot verify implementations that rely on library data structures (e.g., \cref{fig:observer,fig:iterator}).

Another important point of comparison is the level of coupling between collaborating classes,
which we can illustrate using the Master clock example.
In~\cite{Dynamic}, class \e{MASTER} requires complete knowledge of the invariant of class \e{CLOCK}, 
which breaks information hiding (in particular, \e{MASTER} has to be re-verified when the invariant of \e{CLOCK} changes).
The update guards of~\cite{Friends} can be used to declare that slaves need not be notified as long their master's time is increased;
this provides abstraction over the slave clock's invariant,
but class \e{MASTER} still depends on class \e{CLOCK}---where the update guard is defined.
In general, the syntactic rules of~\cite{Friends} require that subject classes declare all potential observer classes as ``friends''.
In SC, update guards are defined in subject classes; 
thus we can prove that \e{tick} maintains the invariants of all observers without knowing their type.
Among the other approaches, only history invariants~\cite{History} support the same level of decoupling, but they cannot preserve stability with the \e{reset} method.

\ifextended
\fakepar{Reasoning without invariants}
Other, more fundamental verification methodologies not based on invariants, such as dynamic frames~\cite{DynFrames} and separation logic~\cite{Reynolds02}, can fully handle all the six benchmark problems.
The generality they achieve is, however, not without costs, as one loses stability of consistency properties (e.g., \e{SUBJECT} is not required to notify all its observers).
Using recursive predicates instead of invariants to define global consistency also loses locality of specifications: for example, updates local to a node in a doubly-linked list require to reason about the whole list; and one node that becomes inconsistent during global updates in the Composite example makes the whole structure inconsistent (instead of just the parent).
Recursive predicates over cyclic structures such as PIP also introduce non-trivial proof obligations to check they are well-founded.

\fakepar{SAVCBS workshops solutions}
SC also fares favorably compared against the solutions submitted to the SAVCBS workshops~\cite{SAVCBS} challenges (Iterator, Observer, and Composite).
Considering only solutions for general-purpose languages and targeting complete requirement specifications, there are two solutions to the Iterator problem and two to the Composite problem.
One solution to the Iterator uses JML and ESC/Java2; the collaborating parts of the invariants are, however, described by pre- and postconditions.
One solution to the Composite also uses JML; it is hard to compare it to other solutions as it is based on model programs and proves invariant preservation only for methods that refine the model program used as specification.
One solution to the Composite uses separation logic and VeriFast; the specification overhead for clients is higher than in our solution but there is no ghost state in the nodes (which has to be updated during global modifications), thus it has advantages and disadvantages compared to our solution.
\caf{Anything about the separation logic solution to the Iterator?}
\fi

\ifextended
\section{Conclusions and Future Work}
\label{sec:conclusions}

We presented \emph{semantic collaboration}: a novel methodology for specifying and verifying invariants of arbitrary object structures.
Compared to existing invariant protocols, it offers considerable flexibility and conceptual simplicity, 
as it introduces no ad hoc syntax and does not syntactically restrict the form of invariants.
We implemented semantic collaboration as part of the AutoProof Eiffel program verifier.
Our experiments with six challenge problems demonstrate the wide applicability of the methodology.
\fi

In an ongoing effort, we have been using SC to verify a realistic data structure library.
This poses new challenges to the verification methodology; in particular dealing with \emph{inheritance}.
Rather than imposing severe restrictions on how invariants can be strengthened in subclasses,
we prefer to re-verify most inherited methods
to make sure they still properly re-establish the invariant before wrapping the \e{Current} object.
We maintain that this approach achieves a reasonable trade-off.

\ifextended
When it comes to reasoning about invariants, sequential and concurrent programs each have their distinctive challenges.
In a sequential setting, one typically performs state updates in series of steps that temporarily break object consistency; this is acceptable since intermediate states are not visible to other objects.
A sequential invariant protocol must adequately support such update schemes, while making sure that invariants hold at ``crucial'' points.
Concurrent invariant protocols deal with different schemes, and hence have different goals.
For this reason, we do not recommend extending SC to deal with concurrent programs; rather, it could be \emph{combined} with an invariant protocol for concurrent programs,
as done in VCC~\cite{VCC}.
\fi

\bibliographystyle{splncs03}
\bibliography{semicola}

\end{document}